\documentclass{sf2a-conf2012}
\usepackage{graphicx}
\usepackage{hyperref}
\usepackage[]{natbib}  
\usepackage[cyr]{aeguill}
\usepackage{epstopdf}

\def\BibTeX{{\rm B\kern-.05em{\sc i\kern-.025em b}\kern-.08em
    T\kern-.1667em\lower.7ex\hbox{E}\kern-.125emX}}
\bibpunct{(}{)}{;}{a}{}{,}  

\begin{document}

\TitreGlobal{SF2A 2012}


\title{Status and results from the RAVE survey}

\runningtitle{RAVE science review}

\author{A. Siebert}\address{Observatoire Astronomique, Universit\'e de
  Strasbourg, CNRS, 11 rue de l'universit\'e, 67000 Strasbourg, France}

\author{the RAVE collaboration}

\setcounter{page}{237}


\maketitle


\begin{abstract}
  RAVE,  the RAdial  Velocity Experiment,  is a  large  spectroscopic survey
  which collects  spectroscopic data for  stars in the  southern hemisphere.
  RAVE uses the AAO Schmidt  telescope with a wavelength coverage similar to
  Gaia but a  lower resolution of R=7,500.  Since  2003, RAVE collected over
  500,000 spectra providing an  unprecedented dataset to study the structure
  and  kinematics of the  Milky Way  and its  stellar populations.   In this
  review,  we  will summarize  the  main  results  obtained using  the  RAVE
  catalogues.
\end{abstract}



\section{The RAdial Velocity Experiment: overview and current status}
  
Understanding the  formation and  evolution of galaxies  is one of  the main
challenge of present day astronomy, and due to our location in its disc, the
Milky Way offers a large amount  of possibility to gain detailed insights on
this  subject.   Progress  on  this  topic  more  and  more  relies  on  the
measurement  of  the  six  dimensions  of  the  phase-space,  positions  and
velocities, and the measurement of  precise chemical abundances for stars in
the  Galaxy.  The measurement  of six  dimensional phase-space  requires the
knowledge of the generally missing line-of-sight (LOS) velocity
which is the primary goal of RAVE.\\

\begin{figure}[ht!]
 \centering
 \includegraphics[width=0.8\textwidth,clip]{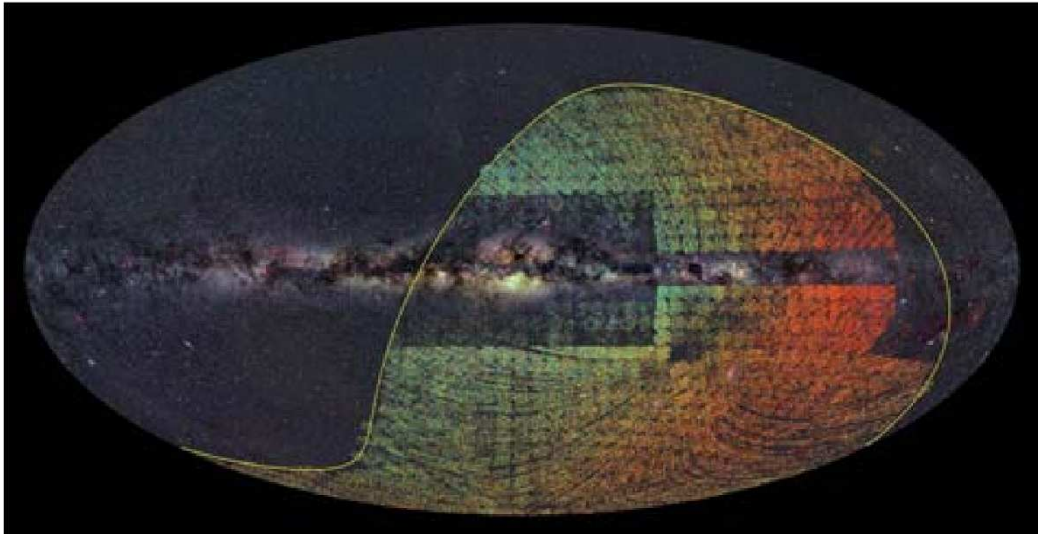}      
 \caption{Aitoff projection  of the RAVE  targets measured as of  June 2010.
   The yellow line  marks the location of the  celestial equator. The colour
   follows  the  heliocentric   line-of-sight  velocity,  red  indicating  a
   velocity  larger than  50~$\mathrm{km}\,\mathrm{s}^{-1}$, blue  lower than
   $-50\mathrm{km}\,\mathrm{s}^{-1}$. The apparent dipole reflects the motion
   of the Sun with respect to the local standard of rest.}
  \label{f:aitoff}
\end{figure}

Taking  advantage  of multi-object  spectroscopy  which  enables to  acquire
spectra for multiple stars  simultaneously, RAVE started its observations in
2003. RAVE uses  the 6dF instrument mounted on the  Schmidt telescope of the
AAO  in Siding  Spring, Australia.   This instrument  enables us  to collect
spectra for up to 150 stars in a 5.8 degrees in diameter field with a single
observation.  The  targeted spectral region,  $\lambda\lambda 8410-8794\AA$,
contains  the infrared  Calcium triplet  and  is similar  to the  wavelength
domain chosen for Gaia. The  effective resolution of $R\sim7,500$ enables us
to measure radial velocities with  a precision better than 5~$\mathrm{km} \,
\mathrm{s}^{-1}$,   the  mode   of  the   distribution  being   better  than
2~$\mathrm{km} \,  \mathrm{s}^{-1}$.  As of  July 2012, RAVE  collected more
than  556,000   spectra  for  more  than  468,000   individual  stars.   The
distribution on  the sky of the observed  RAVE targets, as of  June 2010, is
shown in Fig.~\ref{f:aitoff} and covers a large fraction of the sky
accessible from the southern hemisphere.

So  far,  RAVE  radial  velocities  and associated  measurements  have  been
released to the  community in three data releases:  DR1 \citep{DR1} provided
LOS  velocities  for 25,000  stars  covering  4,700  square degrees  in  the
southern  hemisphere, DR2 \citep{DR2}  released $\sim82,000$  LOS velocities
and estimates of the  atmospheric parameters for $\sim21,000$stars while DR3
\citep{DR3} expended these numbers  to 80,000 LOS velocities and atmospheric
parameters for 40,000 stars.  These catalogues are supplemented by catalogues
providing distance estimates based on RAVE observations \citep{breddels2010,
  zwitter2010,burnett2011}  which  enable to  estimate  the six  dimensional
phase-space data of the targets and are also supplemented by a catalogue of
chemical abundances by \cite{boeche2011}.

The three  first data  releases are  based on an  input catalog  build using
Tycho-2 and  the Supercosmos Sky Survey  (SSS).  The next  data release, DR4
(Kordopatis et al., in prep.), will release data using a new input catalogue
based  on DENIS  I-band  magnitudes and  will  provide improved  atmospheric
parameters using the DEGAS and MATISSE algorithms \citep{K2011,DEGAS,MATISSE}.

In the next section we will review the contribution of RAVE to the study of
the Milky Way and its stellar populations. 
  
\section{Scientific contribution of RAVE}

The depth  and spatial coverage  of the RAVE  survey makes it a  well suited
tool  to study  the  structure and  stellar  populations of  the Milky  Way.
Indeed, the bulk of RAVE  stars samples the Galaxy at distances intermediate
between Hipparcos (the  solar neighbourhood) and the galactic  halo which is
well covered  by the  SDSS.  Although  RAVE was designed  to search  for the
signatures of  the hierarchical  build-up of the  Milky Way, its  design and
coverage proved to  be useful to study many aspects  of the Galaxy.  Indeed,
this region, between  0.5 and 2 kpc  from the Sun, is dominated  by the thin
disc with  an increased  contribution of the  thick disc.  In  the following
presentation of RAVE  results we will try to group  the results according to
some general  topics: structure and kinematics  of the Milky  Way, origin of
the  thin and  thick discs,  moving  groups, accretion  events and  peculiar
objects.
 
\subsection{Structure and  kinematics}

The  first RAVE  catalogue was  released in  2006 and  the  first scientific
result using the  RAVE data followed shortly after  with a new determination
of the  local escape velocity of  the Milky Way  by \cite{smith2007}.  Their
findings of  $v_{esc}=544\,\mathrm{km} \, \mathrm{s}^{-1}$ ($498<v_{esc}<608
\, \mathrm{km}  \, \mathrm{s}^{-1}$  at the 90  per cent  confidence level),
based on  a sample of high-velocity  stars from RAVE  combined to previously
published  data, demonstrates  the  presence of  a  dark halo  in the  Milky
Way. Furthermore, assuming a contracted  NFW halo model, this result implied
a   virial  mass   of   $1.42^{+1.14}_{-0.54}  10^{12}\,\mathrm{M}_{\odot}$,
significantly higher than previous estimates.

Then  \cite{siebert2008},  using a  sub-sample  of  red  clump giants  whose
distance  can   be  estimated  from  the  photometry   alone,  measured  the
inclination of  the velocity  ellipsoid at 1~kpc  below the  Galactic plane.
This inclination, or  tilt, is directly linked to the  shape of the Galactic
potential and of the dark  halo. This measurement (7.3 degrees) was compared
to predictions varying the flattening of  the dark halo and using the latest
mass models  of the  Milky Way  by \cite{BT08}. Although  no model  could be
clearly ruled  out, a nearly spherical  halo at the Sun's  distance from the
Galactic center is preferred.  This also implies the scale length of the
disc to  be in the range 2.5-2.7 kpc.

Focusing   on   the   disc,   its   vertical  structure   was   studied   by
\cite{veltz2008}. In this  work, the authors combined a  subsample of RAVE G
and K type  stars towards the Galactic pole to 2MASS  stars counts and UCAC2
proper motions. They were able  to identify discontinuities in the kinematic
and magnitude  counts, discontinuities  that separate the  different stellar
components.   The  clear kinematic  gap  between  the  thin and  thick  disc
reinforced the view that the thick  disc is unlikely to have formed from the
thin disc in a continuous process.  This work also provided new measurements
of the scale heights of the thin and thick disc, the thin disc scale height
being measured to be $225\pm,10$ pc, and $1048\pm36$ pc for the thick disc.

\cite{casetti2011} combined a  sub sample of 4400 red  clump giants from the
RAVE DR2 catalogue with the  SPM4 \citep{spm4} proper motions to analyse the
three-dimensional  kinematics  of the  thick  disc  population. This  sample
covers distances from 5 to 10 kpc from the Galactic center and reaches 3 kpc
in  height from  the Galactic  plane. They  determined the  global kinematic
parameters  of   the  thick  disc  to   be  $(\sigma_{V_R},  \sigma_{V_{θ}},
\sigma_{V_z})| _{z=1} = (70.4,  48.0, 36.2) \pm (4.1,8.3,4.0) \mathrm{km} \,
\mathrm{s}^{-1}$, with  a tilt  angle of 8  degrees.
This   latter   value   is   in   agreement  with   the   determination   by
\cite{siebert2008} and implies a disc scale length of 2~kpc.

Also, for a good understanding of  the Milky Way, a precise knowledge of the
Sun  velocity  vector  with  respect  to  the  local  standard  of  rest  is
important. This problem  was studied by \cite{veltz2008} using  his model of
the     vertical     structure      of     the     Galactic     disc
and by  \cite{coskunoglu2011}.  This latter  work uses a sub-sample  of RAVE
stars  restricted to  600 pc  from  the Sun,  based on  the photometric  and
spectroscopic properties of the stars.  Their findings are in good agreement
with recent determinations
\citep[see table 1 of][for a summary]{coskunoglu2011}.\\

If star counts and velocity  distributions are useful to decipher the global
properties  of the  Milky Way,  detailed measurements  of the  structure and
kinematics   of  the  stellar   populations  need   the  knowledge   of  the
distance. Distances not only allow us to recover the full velocity vector of
a  star, they  also  allow us  to  sample different  regions  of the  Galaxy
\citep[see  for example]{siebert2008,casetti2011}. If  red clump  stars have
proved to be a precious tool  for such investigations, they represent only a
small fraction of  the RAVE catalogues.  Thanks to  the estimates of stellar
atmospheric parameters,  spectrophotometric distances could  be computed for
most     of    the     RAVE    stars     by    three     different    groups
\citep{breddels2010,zwitter2010,burnett2011}  with   an  overall  very  good
agreement  between  the  groups  while  the  techniques  rely  on  different
assumptions.

The availability  of distances  allow a more  detailed analysis of  the fine
structures in  the disc but also  allows to reduce the  uncertainties on the
global  parameters. For  example,  \cite{karatas2012} revised  the thin  and
thick  discs velocity ellipsoid  measurements using  the \cite{breddels2010}
distances and found an  overall good agreement with previous determinations.
\cite{siebert2011} used  the \cite{zwitter2010} distance  estimates to study
the  mean  velocity  field in  the  Galactic  plane  within 2~kpc  from  the
Sun. They found  a radial velocity gradient whose  origin lies presumably in
non-axisymmetric  perturbations of  the  disc. Assuming  the  local disc  is
mostly perturbed  by spiral arms,  \cite{siebert2012} used the  density wave
model  to model  the observed  velocity field  and constrain  the parameters
describing  the  local  spiral   pattern.   Provided  the  spiral  arms  are
long-lived,  the density  wave  model  with a  2  armed spiral  perturbation
succesfully    reproduces    the    observed    velocity    gradients    and
\cite{siebert2012} estimate  the amplitude of the perturbation  to be 0.55\%
of the background potential, having  a pattern speed of 18.6 $\mathrm{km} \,
\mathrm{s}^{-1}$. This  places the Sun close  to the inner  4:1 resonance, a
location  also suggested  by  the location  of  moving groups  in the  solar
neighbourhood \citep[see for example]{quillen05}.

\subsection{Origin and evolution of the thin and thick discs}

The  origin  and  evolution  of  the  Galactic discs  are  key  elements  in
understanding the  formation of the  Milky Way.  This information  about the
origin is buried  both in the kinematics and in  the chemical composition of
the stars, such as metallicity gradients or eccentricity distributions, both
being available via data provided in  the RAVE catalogues.\\

Focusing on the thick disc chemical properties, \cite{ruchti2010} selected a
sample of 234 metal poor giants  in the RAVE catalogue for a follow-up study
using high-resolution  spectroscopy. A  detailed abundance analysis  of four
$\alpha$  elements  and  iron  abundances  revealed an  enhancement  of  the
$[\alpha/\mathrm{Fe}]$  ratios  as  well  as  a lack  of  scatter  of  these
ratios. This  implies an enrichment  that proceeded by  purely core-collapse
supernovae as well  as a good mixing of the  interstellar medium (ISM) prior
to star formation.  Also the  ratios indicate a similar massive star initial
mass function of the metal poor thick  disc and of the halo.  This leads the
authors to  conclude that  direct accretion of  a dwarf galaxy  with similar
properties than the surviving dwarf galaxies today did not play an important
role in the formation of the thick disc population.

\cite{ruchti2011a}  furthered this  work adding  74 main  sequence  stars to
their sample of giant stars and confirmed their previous result. In addition
they   could   investigate   for    the   first   time   the   gradient   in
$\alpha$-enhancement in  the metal poor  thick disc, finding a  very shallow
gradient ($\frac{\partial  [\alpha/\mathrm{Fe}]}{\partial R,z}<0.03 \pm 0.02
\mathrm{dex} \,  \mathrm{kpc}^{-1}$ for $[\mathrm{Fe/H}]<-1.2 \mathrm{dex}$)
while they find a $+0.01  \pm 0.04 \mathrm{dex} \, \mathrm{kpc}^{-1}$ radial
gradient and  a$ -0.09 \pm 0.05 \mathrm{dex}  \, \mathrm{kpc}^{-1}$ vertical
gradient in iron abundance. This further indicates a good mixing of the ISM
prior to star formation for this population.

The previous  work focused on the  properties of the  metal-poor thick disc,
similar studies used  thin disc stars to constrain  the observed metallicity
gradient  using  the   RAVE  catalogue  \citep{karatas2012,  coskunoglu2012,
  bilir2012} with  values for the  radial gradient ranging from  $-0.04$ to
$-0.07 \mathrm{dex}\,\mathrm{kpc}^{-1}$. Also,  a dependence with age, older
populations showing a shallower  radial gradient is observed. The comparison
to models of the formation of the Galactic disc suggests a contribution from
stellar migration in the shaping of the disc.

Another  aspect  of  the  disc  formation  relies  on  the  distribution  of
eccentricities.  As  shown by  \cite{sales2009}, different scenarios  of the
thick  disc formation  leave  different signatures  in  the distribution  of
eccentricities  of disc  stars.   This signature  has  been investigated  by
different  groups using  the RAVE  data: \cite{casetti2011}  used  red clump
stars from the  DR2, \cite{karatas2012} did the same  exercice using the DR2
sample  together with  \cite{breddels2010}  distances and  \cite{wilson2011}
used the  full RAVE sample. All  studies favour an in-situ  formation of the
thick disc,  the direct accretion  scenario being in  apparent contradiction
with the observed distributions. The good agreement with the result based on
chemical abundances and metallicity gradients further
confirms the in-situ origin of the thick disc.\\

\subsection{Moving groups}

Since their  discovery \citep{eggen1958,eggen1960}, moving  groups have been
the subject of many studies. If  some of the moving groups can be associated
to disrupted  clusters, some of  them are of  resonant origin and it  is now
well  established that  the location  in  velocity space  of these  resonant
structures bear useful informations on the perturbations taking place in the
Galactic
disc.\\

A first attempt to study the moving  groups using the RAVE DR1 data was done
by \cite{klement2008} and continued in  a later work using the DR2 catalogue
\citep{klement2011}.    In  these   works,  the   authors   identified  four
phase-space overdensities,  three of which  were previously known.   The new
stream candidate is on a radial  orbit, suggesting an origin external to the
Milky Way.  However,  their later work using DR2 data  showed that only five
stars belong to that overdensity, preventing clear conclusions to be drawn
at this point on the origin of this overdensity.

\cite{kiss2011} searched the  RAVE database for new members  of young nearby
moving  groups, combining  the  RAVE  data to  stellar  age diagnostics  and
high-resolution optical  spectroscopy follow-up. They were able  to find one
new and five likely members of the $\beta$ Pictoris moving group, one likely
member of  the $\epsilon$ Cha group  and two stars  in the Tucana-Horologium
association, showing the potential of RAVE to increase the census of young
moving groups in the solar neighbourhood.

\cite{hahn2011} combined data from RAVE  and the Sloan Digital Sky Survey to
extract a  sample of stars  within 200~pc of  the Sun. They showed  that the
velocity space structures  seen in the Hipparcos sample  are also present in
these data. They also could associate the Hyades stream to scattering
process at a Lindblad resonance, indicating a resonant origin for this feature.

Thanks  to  the distance  estimates  mentionned  above,  it is  possible  to
reconstruct  the  six-dimensional  phase  space information  and  study  the
evolution   of   the  moving   groups   beyond   the  solar   neighbourhood.
\cite{antoja2012} used the full RAVE sample together with distances and used
a wavelet  analysis to detect the  moving groups. They showed  that the main
groups  observed  in  the  solar  neighbourhood are  large  scale  features,
surviving at  least 1 kpc  from the Sun  in the anti-rotation  direction and
below  the Galactic  plane. Furthermore,  the location  of  these structures
appears to  shift in  the velocity plane  as one  moves away from  the Sun's
location.  These trends are consistent  with dynamical models of the effects
of the  bar and spiral arms, again  indicating a resonant origin  of some of
the moving groups.

\subsection{Signature of accretion events}

One of the  main driver of the  RAVE survey is the search  for signatures of
the hierarchical build-up  of the Milky Way. Although  most of the accretion
events    are    observed   in    the    distant    halo   \citep[see    for
example][]{belokurov2006}, some  are believed to  leave traces in  the inner
parts of galaxies including galactic discs. Indeed, the early simulations of
the  disruption   of  the  Sagittarius  dwarf  galaxy   predicted  that  the
Sagittarius stream could cross the  solar neighbourhood. Such an orbit would
leave an asymmetry in the radial velocity distributions between the northern
and southern  hemisphere.  \cite{seabroke2008} analysed  the RAVE data  in a
cylinder across the disc centered on the Sun and combined this sample to the
local  surveys  from the  Hipparcos  satellite \citep{gcs,famaey2005}.   The
symmetry of the  velocity distributions permits to rule  out the presence of
the Sagittarius stream or the  Virgo overdensity in the solar neighbourhood.
Later simulations of  the disruption of the Sagittarius  dwarf galaxy showed
that the  stream does  not cross the  solar neighbourhood,  intersecting the
Galactic plane further out from the Sun's location, confirming this result.

More  recently, \cite{williams2011}  detected an  overdensity of  stars, the
Aquarius  stream, in  $30^\circ <  \ell  < 75^\circ$  and $-70^\circ  < b  <
-50^\circ$,  with  heliocentric  line-of-sight velocities  $\mathrm{V}_{los}
\sim -200 \mathrm{km} \, \mathrm{s}^{-1}$. These stars are clear outliers in
the  radial  velocity  distribution  and the  overdensity  is  statistically
significant. Analysis  of the RAVE  stars suggest a  metal poor, 10  Gyr old
population.  Using  numerical simulations, they  showed that this  stream is
dynamically  young and  therefore a  debris of  either a  recently disrupted
dwarf galaxy or globular  cluster. High resolution follow-up spectroscopy of
the overdensity members by \cite{wylie2012} showed very little dispersion in
metallicity  (0.1  dex)  indicating  a chemicaly  coherent  structure.   The
location in the nitrogen and  sodium abundances plane further indicates that
the Aquarius stream originates from a disrupted globular cluster.

\subsection{Peculiar objetcs}

The observing strategy of RAVE, a random sampling in magnitude intervals and
no colour selection to mimic a  magnitude limited survey, enables RAVE to be
unbiased with respect  to kinematic selection effects.  If  this strategy is
well suited to the main goals of RAVE, it also preserves the discovery
potential of RAVE.\\

\cite{munari2009}  paper is  a  good example  of  this discovery  potential.
Mining the database, the authors discovered stars in the multi epoch spectra
of RAVE whose radial velocities  and spectra appear dubious for normal Milky
Way objects.   These stars do not belong  to the Milky Way  but are Luminous
Blue   Variable   stars  (LBV)   part   of   the   Large  Magellanic   Cloud
(LMC).  Additional specific  follow-up exposures  were then  taken  one year
appart  for seven LBVs,  including fainter  known LBVs,  to obtain  a fairly
complete sample of LBVs in the  LMC.  Thanks to the multi epoch spectra, the
wind outflow  and variability could be  investigated in some  cases and even
cool companions could be detected.

\cite{ruchti2011b} identified  five lithium rich  field giants in  his metal
poor sample  of RAVE  stars, RAVE being  quite rich  in metal poor  stars as
shown  by  \cite{fulbright2010}.  This  represents  the  largest sample  of
Li-rich giants to date, these objects being rare and important to understand
the structure  and physical processes  taking place in stellar  interior.  A
detailed investigation  of the chemical  abundances by the  authors suggests
that Lithium enrichement in these stars  is due to cool bottom processing, a
different mechanism than the one taking place at the RGB bump.

Among peculiar objects,  binary stars are not uncommon  and the knowledge of
the fraction of stars in binary or multiple systems is an important input of
Galactic models.   In this respect,  identifying multiple stars in  the RAVE
database is important.   Using a method relying on  the properties and shape
of  the  cross-correlation   function,  \cite{matijevic2010}  were  able  to
identify 123 double-lined binary candidates (SB2) in the second data release
of  RAVE, only  eight of  which  were previously  known as  binary stars  in
Simbad. This method is sensitive to systems with orbital periods of 1 day up
to  1  year.   In  a  following paper,  \cite{matijevic2011}  used  repeated
observations  of  RAVE  stars  to identify  single-lined  binary  candidates
(SB1). In this sample of $\sim20000$ stars observed more than once, about 10
to 15\%  of the  stars are detected  as binaries.  Because of the  time span
between observations, the detection is biased towards short periods (days to
weeks).  Therefore the binary fraction reported is a lower limit to the true
binary fraction which is the important quantity for Galactic models.

If   the  analysis   of   the  cross-correlation   functions   and  of   the
re-observations  are  efficient  tools  to detect  binary  stars,  automated
classification of RAVE spectra shows a remarkable ability to detect peculiar
objects.  \cite{matijevic2012} used a local linear embedding technique (LLE)
to automatically classify 350,000 RAVE spectra. If 90 to 95\% of the spectra
belong to normal  single stars, there is a  significant fraction of peculiar
stars  populated   by  the   different  types  of   spectroscopic  binaries,
chromospherically  active stars (both of  them containing  several thousand
spectra) or  other peculiar  objects. Among these  peculiar objects  one can
note  TiO band stars,  carbon stars,  Wolf-Rayet stars,  Be stars  etc. This
shows the large potential of RAVE to increase the statistics and further the
understanding of these rare objects.

Finally, if RAVE observing strategy away from the Galactic plane is meant to
reproduce  the characteristics of  a magnitude  limited sample,  some fields
were  observed in  the Galactic  plane for  calibration purpose  or specific
projects. The study  of Diffuse Interstellar Bands (DIB)  falls in this last
category. \cite{munari2008}  investigated the behaviour of five  DIBs in the
RAVE spectra.  They  could confirm the presence of a  DIB at 8648\AA\, whose
intensity appears unrelated to reddening.  The two DIB at 8531 and 8572\AA\,
appear to be  artifacts due to blends of underlying  stellar lines while the
DIB at 8439\AA\, could not be  resolved due to the strong underlying Paschen
line. However the DIB at 8620\AA\, appears strong and clean in the RAVE
spectra and turns out to be a reliable estimator of reddening.\\

\section{Conclusions}

RAVE operations  started in 2003 and  collected over half  a million spectra
since its  first light.  So  far data were  released to the public  in three
data releases  and complemented by catalogues  containing distance estimates
and  chemical abundances.  The  fourth  data release  is  scheduled in  late
2012/early 2013.\\

If RAVE primary goal is to  search for traces of the hierachical build-up of
the Milky  Way, the design  of the survey  does not restrict  the scientific
capabilities of  the survey. Indeed,  RAVE observations contributed  to many
topics in  Galactic astronomy and  provided useful measurements  and results
that can be grouped in five general topics:
\begin{itemize}
\item structure and  kinematics of the Milky Way and  stellar populations,
\item formation and evolution  of the Galactic discs,
\item velocity space substructures or moving groups,
\item signature of accretion events,
\item search for peculiar objects.
\end{itemize}

If  the  data  collection  is  close  to  completion,  the  vast  amount  of
information  buried in the  RAVE catalogues  and spectra  still has  a large
potential for  being used by the  community.  In this  respect comparison to
models  of the Galaxy,  such as  the Besan\c con  model \citep{besac}  or the
Galaxia  model \citep{sharma2011},  will  be useful  to  interpret the  RAVE
data. This  makes RAVE one of  the major tools for  understanding our Galaxy
until the release of the Gaia catalogue.

\begin{acknowledgements}
  Funding  for  RAVE  has  been  provided by:  the  Australian  Astronomical
  Observatory;  the  Leibniz-Institut fuer  Astrophysik  Potsdam (AIP);  the
  Australian  National  University;  the  Australian Research  Council;  the
  French National Research Agency;  the German Research Foundation (SPP 1177
  and SFB  881); the European  Research Council (ERC-StG  240271 Galactica);
  the  Istituto  Nazionale  di  Astrofisica  at Padova;  The  Johns  Hopkins
  University; the National Science  Foundation of the USA (AST-0908326); the
  W. M. Keck foundation;  the Macquarie University; the Netherlands Research
  School  for  Astronomy;  the  Natural Sciences  and  Engineering  Research
  Council  of Canada;  the  Slovenian Research  Agency;  the Swiss  National
  Science Foundation; the Science \& Technology Facilities Council of the UK;
  Opticon;  Strasbourg  Observatory;  and  the  Universities  of  Groningen,
  Heidelberg     and    Sydney.     The    RAVE     web    site     is    at
  \url{http://www.rave-survey.org}
\end{acknowledgements}

\bibliographystyle{aa}  
\bibliography{siebert_SF2A} 

\end{document}